\documentclass[useAMS,usenatbib]{mn2e}
\usepackage{graphics, amsmath, graphicx,color,tabularx}
\renewcommand\sun{\hbox{$\odot$}}

\renewcommand\degr{\hbox{$^\circ$}}

\renewcommand\arcsec{\hbox{$^{\prime\prime}$}}

\def\iraf{\textsc{iraf}}
\def\gemini{\textsc{gemini}}
\def\nifs{\textsc{nifs}}
\def\ccdproc{\textsc{ccdproc}}
\def\kpnoslit{\textsc{kpnoslit}}
\def\molly{\textsc{molly}}

\title[The mass of the black hole in GRS 1915+105]{The mass of the black hole in GRS 1915+105: new constraints from IR spectroscopy}
\author[D. J. Hurley et al.: The mass of the black hole in GRS 1915+105]{D. J. Hurley$^{1}$\thanks{E-mail:
d.j.hurley@mars.ucc.ie}, P. J. Callanan$^{1}$, P. Elebert$^{1}$ and M. T. Reynolds$^{2}$\\
 $^{1}$Department of Physics, University College Cork, Cork, Ireland.\\
$^{2}$Department of Astronomy, University of Michigan, 500 Church St., Ann Arbor, MI 48109\\
}

\begin{document}

\date{Accepted 2012 December 31.  Received 2012 December 25; in original form 2012 March 15}

\pagerange{\pageref{firstpage}--\pageref{lastpage}} \pubyear{2012}

\maketitle

\label{firstpage}

\begin{abstract} 
GRS~1915+105 has the largest mass function of any Galactic black hole
system, although the error is relatively large. Here we present
 spectroscopic analysis of medium-resolution IR VLT archival data of
GRS~1915+105 in the {\it K}-band. We find an updated ephemeris, and
report on attempts to improve the mass function by a refinement of the radial velocity estimate. We show that the spectra are
significantly affected by the presence of phase-dependent CO bandhead
emission, possibly originating from the accretion disc: we discuss the impact
this has on efforts to better constrain the black hole mass. We report on a possible way to measure the radial velocity utilising apparent {\it H}-band atomic absorption features and also discuss the general uncertainty of the system parameters of this well-studied object.
\end{abstract}
\begin{keywords}
 stars: binaries -- infrared: stars -- stars: individual: GRS
 1915+105. \end{keywords}

\section{Introduction}
Since its discovery by \citet{b1} GRS~1915+105 remains
one of
the most intensively studied of all the Galactic X-ray sources, and is the prototypical Galactic `microquasar'.
This system is of particular importance, not only for our understanding
of the formation of jets near black holes, but also because it is
regarded as a Galactic analogue to Active Galactic Nuclei \citep{b7,b2,b4,b5}.
Due to its location only 0.2 arcsec from the Galactic plane, GRS~1915+105
suffers from strong interstellar absorption: the {\it V}-band extinction ($ A_{\rm v}$) is $ \rm 19.6 \pm 1.7$
mag \citep{b58}. For this reason observations in the {\it K}-band
have been particularly critical, leading to the
identification of the secondary \citep{b4}, a radial velocity estimate for the secondary \citep{b5}
and the possible determination of a photometric period \citep{b17}. The radial velocity study yielded a
mass function of $ \rm 9.5 \pm 3.0 $ $\rm M_{\sun} $ which, when combined with inclination estimates \citep{b18,b5}
indicated a primary mass of $ \rm 14.0 \pm 4.0 $ $\rm M_{\sun} $. This not only confirmed the presence of a black hole in the system but indicated it to be one the most massive black holes discovered to date in a Galactic X-ray binary. Other examples of massive primaries include Cygnus X-1 with a primary mass of $ \rm 14.8 \pm 1.0 $ $\rm M_{\sun} $ \citep{b112} and V404 Cyg, which until recently was thought to harbour a $ \rm 12 \pm 2 $ $\rm M_{\sun} $ \citep{b113} but has now been reduced to $ \rm 9.0^{+0.2}_{-0.6} $ $\rm M_{\sun} $ \citep{b114}. \\

Following the estimate by \citet{b5}, \citet{b14} noted that this mass falls outside the theoretical distribution range for our Galaxy as calculated by
\citet{b13}. However, since the error on the black hole mass is so
large ($ \rm \pm 4.0 $ $\rm M_{\sun} $), it is clear that a more accurate estimate is required before we
can make a meaningful comparison to the theoretical mass
distribution. In addition, \citet{b34} derive a lower limit for
the dimensionless spin parameter of $ \rm a_{*} > 0.98 $ making GRS
1915+105 a near maximally spinning Kerr black hole. This conclusion
was drawn from an analysis of RXTE and ASCA data for the
thermal state of GRS~1915+105, and a model of the X-ray continuum of a
fully relativistic accretion disc. As discussed by \citet{b34}, determining the mass of, and distance
to, GRS~1915+105 is of fundamental importance for verifying this
method of spin determination.\\
\begin{center}
\begin{table*}
\begin{center}
 \begin{minipage}{165mm}
  \caption{Complete list of archival data used including total exposure times for each night from 1999-2002.}
\resizebox{16.5cm}{!}{  
\begin{tabular}{@{}lllllllllll@{}}
  \hline
& Date & No. of Exp & Total & Phase &  Date & No. of Exp &  Total & Phase \\
&(UT) &  & (s) & &  (UT) &  & (s) & &\\
    
& 1999 Jul 23 & $\rm 10\times300$s & 3000 & 0.46271  \vline & 2002 Jul 30  & $\rm 10\times300$s & 3000 & 0.23063 \\
& 2000 Apr 23 & $\rm 8\times240$s & 1920 & 0.74271  \vline & 2002 Aug 06  & $\rm 20\times300$s & 6000 & 0.43595 \\
& 2000 May 11 & $\rm 8\times240$s & 1920 & 0.27410  \vline & 2002 Aug 08  & $\rm 10\times300$s & 3000 & 0.49377 \\ 
& 2000 May 23 & $\rm 8\times240$s & 1920 & 0.63179  \vline & 2002 Aug 10  & $\rm 10\times300$s & 3000 & 0.55101 \\ 
& 2000 Jun 10 & $\rm 8\times250$s & 2000 & 0.16171  \vline & 2002 Aug 12  & $\rm 10\times300$s & 3000 & 0.63898 \\ 
& 2000 Jun 13 & $\rm 8\times250$s & 2000 & 0.25325  \vline & 2002 Aug 13  & $\rm 10\times300$s & 3000 & 0.64005 \\ 
& 2000 Jun 18 & $\rm 8\times250$s & 2000 & 0.39898  \vline & 2002 Aug 14  & $\rm 20\times300$s & 6000 & 0.65954 \\  
& 2000 Jul 03 & $\rm 8\times250$s & 2000 & 0.83817  \vline & 2002 Aug 17  & $\rm 20\times300$s & 6000 & 0.75728 \\  
& 2000 Jul 09 & $\rm 8\times250$s & 2000 & 0.01589  \vline & 2002 Aug 19  & $\rm 20\times300$s & 6000 & 0.81673 \\
& 2000 Jul 12 & $\rm 8\times250$s & 2000 & 0.10582  \vline & 2002 Aug 29  & $\rm 20\times300$s & 6000 & 0.11109 \\
& 2000 Jul 14 & $\rm 8\times250$s & 2000 & 0.16342  \vline & 2002 Aug 31  & $\rm 20\times300$s & 6000 & 0.17236 \\ 
& 2000 Jul 17 & $\rm 8\times250$s & 2000 & 0.25120  \vline & 2002 Sep 02  & $\rm 10\times300$s & 3000 & 0.22992 \\ 
& 2000 Jul 23 & $\rm 8\times250$s & 2000 & 0.46333  \vline & 2002 Sep 05  & $\rm 10\times300$s & 3000 & 0.28863 \\ 
& 2000 Jul 27 & $\rm 8\times250$s & 2000 & 0.54704  \vline  \\
& 2000 Aug 01 & $\rm 8\times250$s & 2000 & 0.69426  \vline  \\
& 2000 Aug 21 & $\rm 8\times250$s & 2000 & 0.28594  \vline  \\
\hline
\end{tabular}
}
\end{minipage}
\end{center}
\label{table:data}
\end{table*}
\end{center}

Here we present an analysis of archival Very Large Telescope (VLT)
data and a re-analysis of the data presented by \citealt{b4,b5}. In particular, we show in what follows that, near orbital phase
0.25, the absorption lines are corrupted by emission, which
will have a significant effect on the amplitude of the radial velocity variation. We discuss possible origins for this emission, and the effect it might have on the mass function and primary mass. We also discuss the feasibility of using {\it H}--band spectroscopy to measure the radial velocity of the secondary from narrower atomic features using new Gemini observations.

\vspace{-0.2in}
\section{Observations}
\subsection{ESO VLT archival data}
The European Southern Observatory (ESO) archival data used here were
taken with the Infrared Spectrometer And Array Camera (ISAAC) on the
8-m VLT Antu telescope at Cerro Paranal
(Chile). The short wavelength arm
of ISAAC equipped with a
$1024 \times 1024$ pixel Rockwell HgCdTe
array with an image scale of
0.147 arcsec pixel$ ^{-1}$. Using the
medium resolution grating (1.2 \AA\ pixel$ ^{-1}$ in
the $K$-band)
yields a spectral resolution of $\rm \sim$3000 with a 1 arcsec
slit. Science data consisted of several 250--300 s individual
exposures each night (see Table \ref{table:data} for summary of the data used). We present a re-analysis of 16 nights of observations from 2000 May
to August, which were the focus of \citealt{b4,b5}, and 13
more from 2002 July to September which are
presented here for the first time.\\

The bias subtraction,
flat-fielding and sky subtraction were performed using the \iraf\footnote{IRAF is distributed by the National Optical Astronomy
Observatories, which are operated by the Association of Universities
for Research in Astronomy,  Inc., under cooperative agreement with
the National Science  Foundation}\ \ccdproc\ package. The spectra
were extracted using the \iraf\ \kpnoslit\ package.
Initial wavelength calibration was carried out using arc spectra obtained
before or after each science run. However, cross-correlation of these spectra with the OH sky lines indicated a residual wavelength offset of $ \rm \sim$3--5 $\rm\AA$, presumably due to
flexure in the telescope/spectrograph between the times when science exposures and calibration arc frames
were taken. This wavelength shift was corrected for before the telluric
features were removed.
To correct for telluric absorption the
A0 V star HD 179913 was
observed either before or after the science
exposures each night and
the method as outlined by \citet{b8}
utilized.
\begin{figure}
\includegraphics[angle=0, scale=.475]{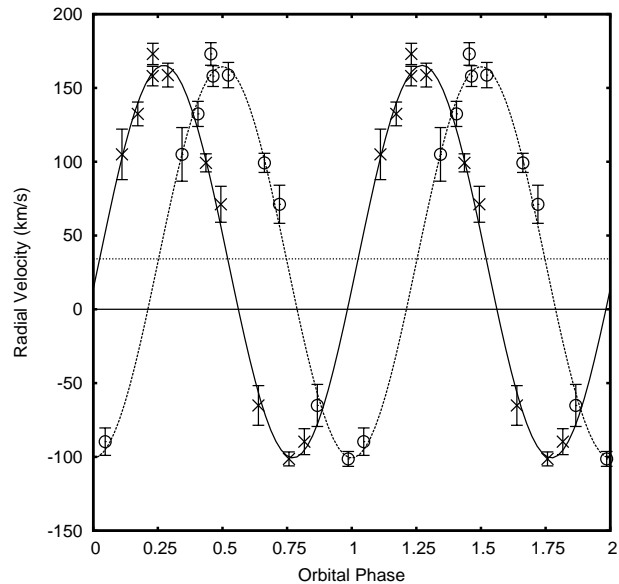}
\caption{Radial velocity curve  of the previously unpublished 2002 dataset folded using the best-fit period of \citet{b5} $ \rm  P_{orb} = 33.5 \pm 1.5$ days (marked with open circles). The fit overlaid has a reduced $ \rm \chi^{2}=1 $. The peak occurs at $ \rm K_{d}=131.5 \pm 2.5$ km $\rm s^{-1}$ with a systemic velocity of $ \rm 33.0 \pm 2.1$ km $\rm s^{-1}$ . As can be clearly seen the phasing of the dataset with the ephemeris of \citet{b4} leads to an offset in the peak from the expected 0.25 to $\rm \sim$0.5. Also plotted is the same data folded on our refined ephemeris (marked with crosses) from the combined dataset (see Figure 3). Note that the refined ephemeris exhibits no significant offset from the expected peak of 0.25.}
\label{fig:rv02}
\end{figure}
\begin{figure*}
\includegraphics[angle=0, scale=0.27]{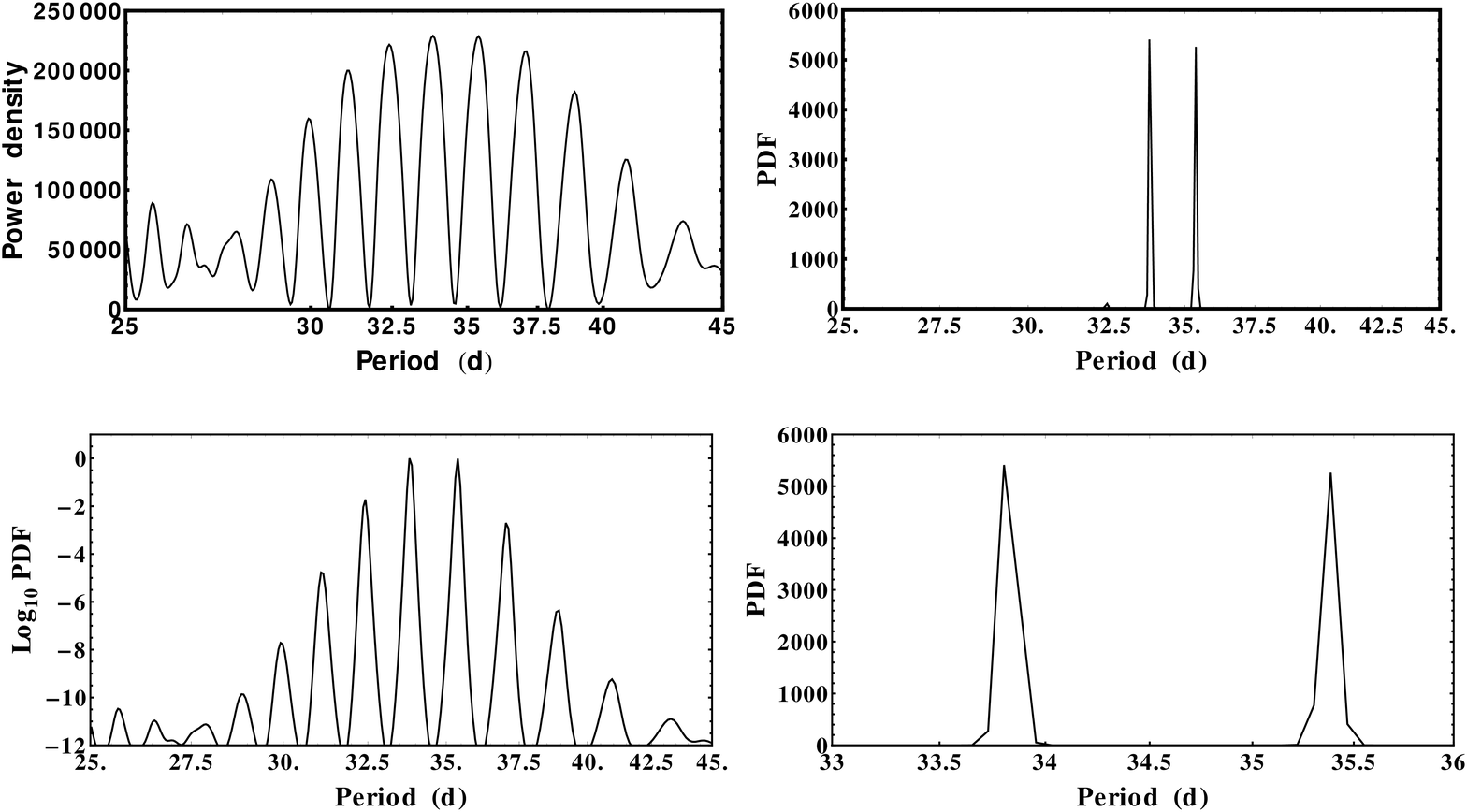}
\caption{(top~left) Lomb-Scargle periodogram of the combined 2000 and 2002 dataset with a $ \rm T_{0} = 2451666.5 $. (top~right) Bretthorst's Bayesian periodogram \citep{b60} which has been marginalized to exclude noise resulting from a non-sinusoidal source. (bottom~left) Normalized Log plot of the PDF to emphasise any lower peaks. (bottom~right) Closer look at the two resulting major peaks which occur at 33.8 and 35.38 days respectively, with each having a FWHM of 0.1 days which we take as the error. Of these only the 33.8 day period successfully phases both datasets correctly.
}
\label{fig:lomb}
\end{figure*}
The data were exported to \molly\footnote{http://www.warwick.ac.uk/go/trmarsh/software} where they were re-binned onto a
common velocity scale. Three techniques were then used to measure the radial
velocity curve: each spectrum of GRS~1915+105 was cross correlated against (i) 
a template spectrum of the KIII standard star HR 8117, (ii) the average of the (remaining) spectra
of GRS~1915+105 and (iii) the average of the spectra occurring around phase 0.75
(see below).  All three techniques resulted in similar velocities within the errors.\\

As a final check, the spectra were also re-extracted using the VLT ISAAC pipeline,
and the resulting radial velocities were again found to be consistent with those determined above once the shift due to instrument flexure had been removed.
\vspace{-0.2in}

\subsection{Gemini-North NIFS data }
In \cite{b4} the presence of atomic absorption features are identified in the {\em H}-band. Motivated by the hope that a radial velocity measurement based on these inherently narrower features would lead to a more accurate and reliable mass function estimate, a total of three 'pathfinder' spectra were obtained using Gemini-North's Near-Infrared Integral Field Spectrometer on the nights of 2010 June 29th, July 6th and July 13th. The observations were scheduled to sample the radial velocity variation at three orbital phases. NIFS provides 3D imaging spectroscopy with a $\rm 3.0\arcsec\times3.0\arcsec$ field of view, equipped with a $\rm 2048 \times 2048 $ Rockwell HAWAII-2RG HgCdTe array with an image scale of $\rm 0.103\arcsec\times 0.043$\arcsec per pixel across and along the slit respectively. Using the H-G5604 grating and the HK-60603 filter yields a spectral resolution of $\rm \sim$5000. Science data consisted of $\rm 5 \times180$s target exposures each night resulting in a S/N $\rm \sim$24. The data were reduced using the \iraf\ \gemini\ \nifs\ package and telluric absorption corrected for by observing the A0V standard HD~182761 before and after science runs. The data were again exported to \molly\ where they were re-binned onto a
common velocity scale. The spectra were then cross correlated against 
the KI III template star HD 83240, taken from the catalogue of high resolution VLT IR spectra of \citet{crires}. A stringent mask of the spectra was used in each case to utilise only the lines identified in Figure \ref{fig:spech} for the cross correlation. \\

\section{Results}
\subsection{ESO VLT archival data}
\begin{figure}
\includegraphics[angle=0, scale=.475]{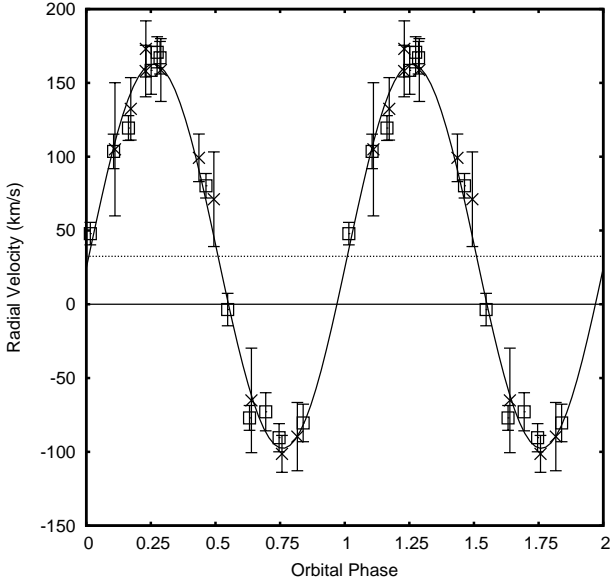}
\caption{Radial velocity curve  of the combined 2000 (marked with open squares) and 2002 (marked with crosses) datasets folded over the best-fit period of  $\rm  P_{orb} = 33.8 \pm 0.1 $ days obtained from the Lomb-Scargle periodogram.  The overlaid fit has a reduced $ \rm \chi^{2}=1 $. The peak occurs at $ \rm K_{d}=131.9 \pm 3.3$ km $\rm s^{-1}$ with a systemic velocity of $ \rm 34.2 \pm 2.5$ km $\rm s^{-1}$. The scaling of error bars to produce a reduced $ \rm \chi^{2}=1 $ makes the 2002 error bars larger than those in Figure 1. This is due to the larger scatter of the points in the combined dataset. }
\label{fig:rv0002}
\end{figure}
In Figure \ref{fig:rv02} we plot the previously unpublished data from 2002 only, folded using the ephemeris of \citet{b5}. The maximum velocity shift is seen to occur at phase 0.5, whereas it should occur at phase 0.25 by definition, suggesting that a refinement in the ephemeris of \citet{b5} is needed. Both the 2000 and 2002 datasets yield radial velocity amplitudes consistent with the result of $ \rm 140 \pm 15$ km $\rm s^{-1}$ found by \citet{b5}, although the systemic velocity ($\rm \gamma$) differs somewhat to the previously published value of $ \rm -3 \pm 10$ km $\rm s^{-1}$.  This difference is most likely due to the skyline correction outlined in Section 2.1 as prior to this correction we find a systemic velocity that is consistent with that of \citet{b5} for the same dataset.\\ 

We next combined both the 2000 and 2002 datasets in order to derive an improved ephemeris by performing a period analysis via a Lomb-Scargle periodogram and then refining the results using Bretthorst's bayesian periodogram \citep{b60} to clean out any false signals, the results of which are shown in Figure \ref{fig:lomb} (calculated using $ \rm T_{0} = 2451666.5$ from \citet{b5}). The peak that corresponds to a period of  $\rm 33.8\pm 0.1$ is the only one to successfully phase all the data consistently and so we adopt this as our revised orbital period for GRS~1915+105. This period is consistent with that of \citet{b5} ($ \rm33.5 \pm 1.5$ days) to within 1--$\rm \sigma$, but does not agree with the low amplitude {\it K}--band modulation discussed by \cite{b17}. Folding all the data on this revised period yields the radial velocity curve shown in Figure \ref{fig:rv0002}. A sinewave fit to this plot yields a secondary radial velocity semi-amplitude ($ \rm {\it K_{d}} $) of $ \rm 131.9 \pm 3.3$ km $\rm s^{-1}$ and a systemic velocity ($\rm \gamma$) of $ \rm 34.2 \pm 2.5$ km $\rm s^{-1}$. To estimate the uncertainties on derived quantities, we used a Monte Carlo analysis. This involved selecting 100000 random samples from distributions of each input parameter {\it i}, $ \rm {\it K_{d}} $, $\rm {\it P_{orb}}$ and $\rm {\it M_{D}}$  (based on
their means and standard deviations), and calculating the results based on the output distributions. Combining our initial estimates in $\rm {\it P_{orb}}$ and $\rm {\it K_{d}}$ with the suggested donor mass ($\rm {\it M_{D}}$) of $\rm 0.8 \pm 0.5$ $\rm M_{\sun} $  \citep{b50} and the binary inclination ({\it i}) of $ \rm 66 \pm 2 $ degrees \citep{b18} the mass function: 
\begin{displaymath}
\rm f(M) = \frac{( {\it M_{c}} sin \it{i} )^{3} }{ ( {\it M_{c}} + {\it M_{D}} )^{2} } = \frac{{{\it P_{orb}}}{{\it K_{d}}}^{3} }{2 \pi G}
\end{displaymath}
is found to be $\rm 8.0 \pm 0.6\  M_{\sun}$ ( at the 68 per cent confidence interval), which is slightly lower than the $ \rm 9.5 \pm 3.0 $ $\rm M_{\sun} $ determined by \citet{b5}, but consistent within their $ \rm 1 \sigma $ error. The mass of the primary is found to be $ \rm 12 \pm 1.4 $ $\rm M_{\sun}$.\\
\begin{figure*}
\includegraphics[angle=-90, scale=0.8]{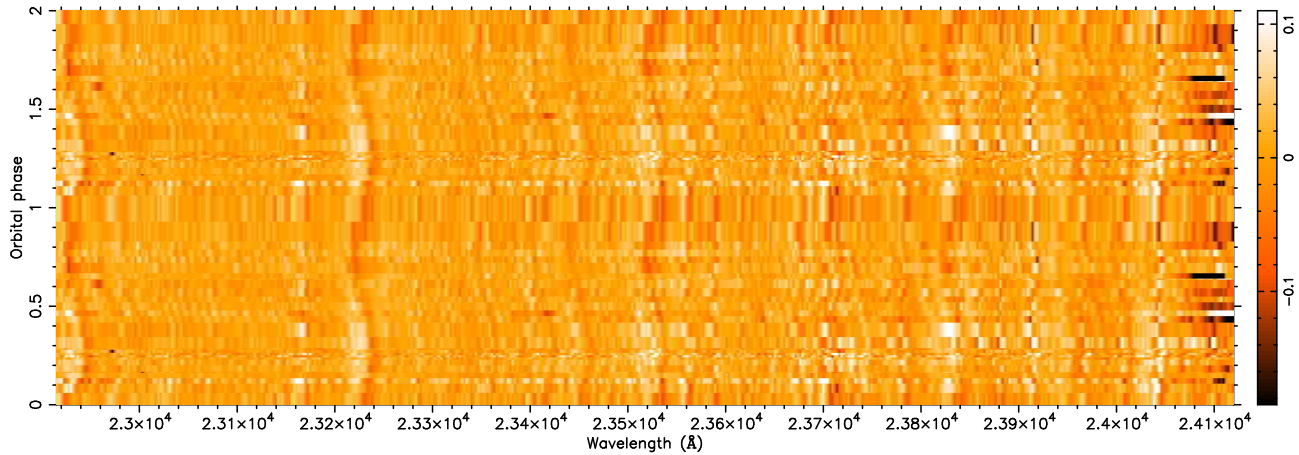}
\caption{Trailed spectrum of all data, repeated twice for clarity. Lighter colours indicate emission and darker absorption. Notice the emission is visible only in the the first half of the orbital cycle, blueward of each bandhead, and is strongest near phase $\sim$0.25 (see electronic version also). Intermittent emission at $\lambda 23160$ $\rm \AA$ is a residual telluric feature.}
\label{fig:sine}
\end{figure*}

\begin{figure}
\includegraphics[angle=-90, scale=.35]{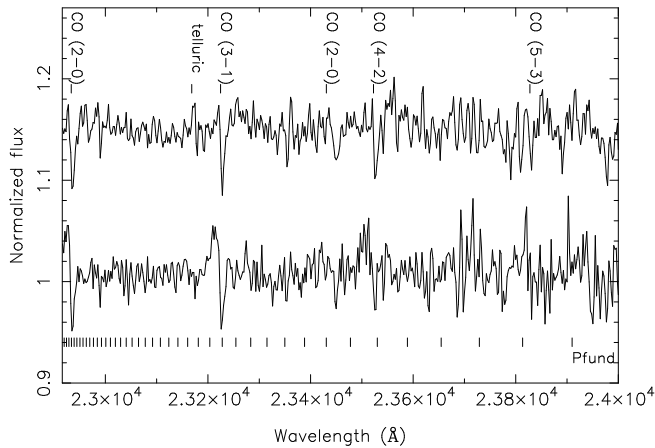}
\caption{Bottom: Average of 2002 spectra between 0.15 and 0.35 in phase. Top: Average of 2002 spectra taken between 0.65 and 0.85. Again, as with the 2000 spectra, the emission is present in the same orbital ranges. Again, the positions of the Pfund lines are indicated for reference.}
\label{fig:aver02}
\end{figure}

However, in Figure \ref{fig:sine} we plot the trailed spectra for the combined dataset phased on our refined ephemeris: emission can be seen in the first half of the orbital range, indicated by the lighter regions, blueward of all the CO bandhead absorption features (this is more apparent in the electronic colour version). The emission rises and falls over the first half of the orbit and is predominantly strongest near phase 0.25. To demonstrate this emission more clearly, in Figures \ref{fig:aver00} \& \ref{fig:aver02} we plot the summed spectra, phased on the revised ephemeris, for both the 2000 and 2002 datasets. In each case the data is further separated into two  distinct orbital phase windows (0.15-0.35 and 0.65-0.85) to illustrate the presence and absence of the emission near 0.25 and 0.75 respectively. As can be clearly seen, there is significant CO bandhead emission present in the first half of the orbital phase range, not in the latter. This emission is present in the two epochs of observation (most noticeably near the $\rm {}^{12}CO$ (2-0) and (3-1) transitions) indicating that it is not just a transient phenomenon. Whereas variability in the bandhead absorption had been reported by previous authors \citep{b50}, the fact that it occurs predominantly near orbital phase 0.25 has important implications for the radial velocity curve, which we discuss in Section 4.\\
\vspace{-0.3in}
\subsection{Gemini-North NIFS data}
\begin{figure*}
\includegraphics[angle=-90, scale=.635]{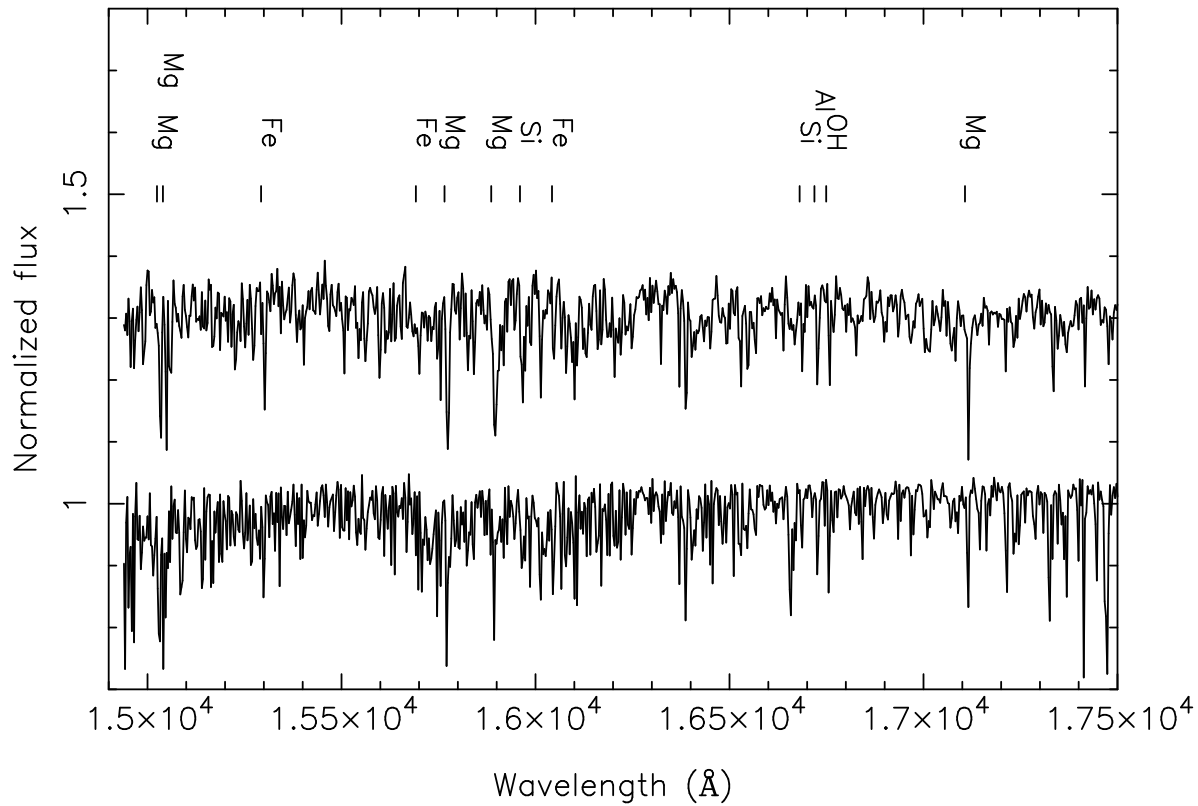}
\caption{{\it H}-Band spectrum confirming the presence of atomic features. Top: A spectrum from a single nights' observation with Gemini. All features indicated are found in each of the three spectra. Equivalent widths of selected lines can be found in Table \ref{table:data2}. Bottom: Template star HD 83240 from the catalogue of \citet{crires}}
\label{fig:spech}
\end{figure*}

\begin{table*}{}
\centering
\resizebox{16.5cm}{!}{
  \begin{tabular}{|l||c|c|c|c|c|}
  \hline
Spectrum & Mg (15770.9) &  Si (1668.53) & Mg (17113.08) & Al (16723.5) & Velocity \\
 & (\AA) & (\AA) & (\AA) &  (\AA) & ($\rm km$ $\rm s^{-1}$)\\ \hline
 2010 Jun 29 & $\rm 2.02 \pm 0.16 $ & $\rm 0.59\pm0.10$ & $\rm 1.07\pm 0.10$ &  $\rm 1.25\pm 0.22$ & $\rm 9.1 \pm 20.5$ \\ 
 2010 Jul 06 & $\rm 2.47 \pm 0.15$ & $\rm 0.44 \pm 0.10$ & $\rm 0.92 \pm0.10$ & $\rm 1.31\pm0.15$ & $\rm 54.4 \pm 21.3$ \\ 
 2010 Jul 13 & $\rm 2.55 \pm 0.15$ & $\rm 0.49 \pm 0.10$ & $\rm 1.05\pm 0.11$ & $\rm 1.07\pm0.13$ & $\rm 62.5 \pm 21.9$ \\ 
 HD 83240 & $\rm 2.23\pm0.20$ & $\rm 0.47\pm 0.10 $& $\rm0.99\pm0.09$ & $\rm 0.567\pm0.07$ & \\
\hline
\end{tabular}
}
\caption{Equivalent widths of {\it H}--band spectra. Only the deepest lines are listed. Also listed are the corresponding radial velocity measurements, and their associated errors, from the cross correlation analysis in MOLLY.}
\label{table:data2}
\end{table*}
From our NIFS spectra (see Figure \ref{fig:spech}) we confirm the presence of many of the atomic features noted by \citet{b4}: namely the Al triplet ( $\lambda 16723.5$ \AA, $\lambda 16755.2$ \AA, $\lambda 16767.9$ \AA), Si I ($\lambda 15964.4$ \AA, $\lambda 16685.3$ \AA), Mg I ($\lambda 15029.1$ \AA, $\lambda 15044.3$ \AA, $\lambda 15770.1$ \AA, $\lambda 17113.3$ \AA), Fe I ($\lambda 15297.3$ \AA, $\lambda 15969.1$ \AA, $\lambda 16047.1$ \AA) and OH ($\lambda 16753.8$ \AA). Equivalent widths of some of the major lines can be found in Table \ref{table:data2}. We do not detect Br series ($\rm\lambda $ or $\rm\eta$), CO or He I. We first used these spectra to independently estimate the effective temperature ($ \rm T_{eff}$) of the secondary, from Equation 3 of \citet{Le}, using the EW of the Mg I ($\lambda 17113.3$ \AA) line. From this  we estimate an $\rm T_{eff}$ of $ \rm 4540^{+505}_{-505} $ K which is reassuringly consistent with the K2 III classification already ascribed by \citet{b4}.\\ 

The resulting velocities from our cross correlation analysis (see Section 2.2 for details) are shown in Table \ref{table:data2}. These were then searched for the expected variation due to the motion of the secondary, based on the results of the {\it K}--band observations outlined in Section 3.1. However, even with the improvement in the error of the ephemeris (Section 3.1), over the intervening 110 cycles between the {\it K } \& {\it H}--band observations the uncertainty in phase information accumulates to $\rm \sim \pm$0.33. Hence, in constraining $\rm {\it K_{d}}$, the absolute phase was left as a free parameter. We used the $\gamma$ velocity found in Section 3.1.
To our surprise, we do not find the radial velocity variation expected, based on the CO band-head measurements discussed in Section 3.1 and \citet{b5}. Our measurements are consistent only with an upper limit of $\sim$50 km s$^{-1}$ on $\rm {\it K_{d}}$, although this depends strongly on the assumed value of $\rm \gamma$. 
\vspace{-0.2in}
\section{Discussion}
\subsection*{H-Band Atomic Absorption}
It is clear that some caution should be used when interpreting the radial velocity variation (or lack thereof) in the {\it H}-band, as we have only three measurements at our disposal. If the lack of variation (compared to what is expected from the {\it K}--band) is real, then we must look for alternative sources of the {\it H}-band emission.
For example, the work of \citet{b56} and \citet{b57} suggests that GRS~1915+105 may possess a circumbinary disc of material: if we assume such a disc is the source of the {\it H}-band absorption features, then we can use the FWHM of the latter to constrain the location of the emission.  The observed FWHM of the absorption features is $\sim$60 km s$^{-1}$, and assuming a Keplerian disc orbiting a binary of total mass 12.8 $\rm M_{\odot}$, we estimate a radius of 530 $\rm R_{\odot}$, well outside the binary radius of 100 $\rm R_{\odot}$.
However, it remains suspicious that the {\em H}-band spectrum, despite the lack of radial velocity variation, so closely matches the expected spectral type of the secondary. It is clear that more systematic observations of these spectral features are warranted. \\
\vspace{-0.7cm}

\subsection*{CO Bandhead Emission}
\begin{figure}
\includegraphics[angle=-90, scale=.35]{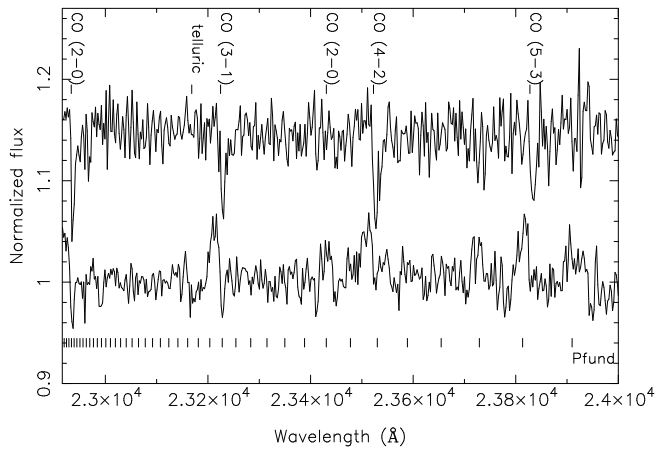}
\caption{Bottom: Average of 2000 spectra between 0.15 and 0.35 in phase. Top: Average of 2000 spectra taken between 0.65 and 0.85. Notice the emission present in the former compared to the latter. The positions of the Pfund lines are indicated to demonstrate that they cannot account for the structure of the observed emission.}
\label{fig:aver00}
\end{figure}

The CO emission is a rare phenomenon in accreting binary systems: to
our knowledge, it has only previously been observed in the well known
cataclysmic variable WZ Sagittae \citep{b10}. As with \citet{b10}, we
also believe that H Pfund emission is unlikely to be a major
contributor to the observed emission. Higher order lines ($ > $ P 33) blend to form a continuum at shorter wavelengths (see Figure \ref{fig:aver00}) and do not exhibit the discrete band structure blueward of each CO bandhead that is observed. Also, the intensity of Pfund emission lines should increase from shorter to longer wavelengths, whereas the emission that we observe is approximately uniform at each site.  Hence we believe that CO bandhead emission is the best explanation for the observed emission features.\\
With a dissociation energy of 11.1 eV, the site of this emission must
reside in a relatively cool, outer part of the disc. Such a region
cannot be directly illuminated by X-rays from the inner disc, and must instead be shielded from it. This could be provided,
for example, by a warped accretion disc, which in turn has been
invoked to explain the long term periodicities in the X-ray light curve
of GRS~1915+105 \citep{b47}. \\

However, it is the fact that the emission occurs preferentially near
orbital phase 0.25 that most interests us here. Such emission may
become more visible near this phase because the absorption lines of
the secondary are at their maximum redshift at this phase, making
emission from the accretion disc easier to observe. However, the 
same should be true at maximum blue shift (phase 0.75), which is not the case for our data.\\

Alternatively, it may simply be that the shielded region of the disc
is best visible near binary phase 0.25 - although, in the context of a warped and
precessing disc \citep{b57}, it is not clear why this should be the case for both
epochs of observations.
Another possible cause for the appearance of the blueshifted emission phenomenon is P-Cygni type emission originating from a wind driven outflow of material, such as proposed by \cite{b57} for GRS~1915+105. This would imply a shell of material somewhere between 200 - 400 $\rm R_{\sun}$ from the primary (\citealp{b115}, Equation 1). The dominant source of uncertainty for this estimate, however, is due to the uncertainty in the distance to the binary.\\

The uncertainty in the distance to GRS~1915+105 also plays a significant role in any attempt to tightly constrain the black hole mass. This is because the estimate for the primary mass stated in Section 3.1 relies heavily on the inclination of the system. While cited by many authors as $\rm 66 \pm 2 \degr $ \citep{b34,van_2010,rahoui_2010}, this estimate is based in turn on an accurately known distance. However, the distance to GRS~1915+105 is still a matter of some debate;  the conservative estimate of $\rm 9.0 \pm 3.0$ kpc as proposed by \citet{b58} would suggest a binary inclination of $ \rm 58 \pm 11 \degr $ (using Equation 4 from \citealp{b18}). Combining this with a donor mass of $\rm 0.8 \pm 0.5$ $\rm M_{\sun} $  \citep{b50}, yields a considerably more uncertain primary mass of $ \rm 16.7 \pm
7.4 $ $\rm M_{\sun}$.\\

To estimate the distance to GRS~1915+105, \citet{b5} combined their estimate of $\rm \gamma$ with the Galactic rotation curve of \citet{fich} to yield $\rm D=12.1\pm0.8$~kpc. Using the same technique, our estimate of $\rm \gamma$ (see Section 3.1) yields $\rm D=9.4\pm0.2$~kpc, where the low error is due to the fact that we assume a flat rotation curve and thereby only the uncertainty on $\rm \gamma$ contributes. This in turn would suggest an inclination of $ \rm 62 \pm 2 \degr $ (again from \citet{b18}), and a primary mass of  $ \rm 12.9 \pm 2.4 $ $\rm M_{\sun}$.\\

Finally, it is also clear that the black hole mass estimate will also be affected by the
presence of the bandhead emission. Specifically, near phase 0.25, the
absorption line measurements will be skewed to higher velocities,
artificially increasing the amplitude of the radial velocity curve. In addition, the presence of the bandhead emission will also affect the measurement of $\rm \gamma$ in the same manner. Hence, in this sense, the mass function we have derived must be considered an upper limit only.  
Further attempts to
refine the mass function and $\rm \gamma$ of GRS~1915+105 utilizing {\em K}-band spectroscopy will depend on the feasibility
of decontaminating the absorption features from this emission
contribution, especially near phase 0.25. This will require higher resolution and S/N spectra than have been obtained thus far.

\vspace{-0.27in}
\section{Conclusions}
We have confirmed the presence of narrow atomic features in the {\it H}-band as noted by \citet{b4}, but our analysis does not provide evidence for the expected radial velocity variation. This may suggest that the origin of the absorption is distinct from the secondary and outside of the binary but more observations are clearly required to confirm this.  Alternatively, the CO bandhead emission observed specifically at phase 0.25 brings into question the reliability of the radial velocity amplitude of the secondary based on the {\it K}-band observations. Higher resolution {\em K}-band observations are required to disentangle the emission from the bandhead absorption so that a reliable radial velocity estimate can be made. 
\vspace{-.32in}
\section*{Acknowledgements}{}

We gratefully acknowledge the use of the MOLLY software package developed by T. Marsh. We would also like to thank the anonymous referee for useful suggestions leading to an improved paper.\\


\label{lastpage}

\end{document}